\documentclass[aps, pre, reprint, superscriptaddress, showpacs, floatfix]{revtex4-1}

\usepackage{graphicx, times, multirow, array, hyperref, color, amsmath}
\usepackage[separate-uncertainty=true]{siunitx}
\usepackage[english]{babel}

\DeclareSIUnit\molar{M}
\newcolumntype{P}[1]{>{\centering\arraybackslash}p{#1}}

\graphicspath{{Figures/}}

\begin{document}

\title{Local details versus effective medium approximation: A study of diffusion in microfluidic random networks made from Voronoi tessellations}
\author{Washington Ponce}
\affiliation{Departamento de Ciencias Exactas, Universidad de las Fuerzas Armadas ESPE - Extensi\'on Santo Domingo. V\'ia Santo Domingo - Quevedo km. 24. Santo Domingo, Ecuador}
\author{Mar\'ia Luisa Cordero}
\email{mcordero@ing.uchile.cl}
\affiliation{Departamento de F\'isica, Facultad de Ciencias F\'isicas y Matem\'aticas, Universidad de Chile, Av. Blanco Encalada 2008, Santiago, Chile}


\begin{abstract}
We measured the effective diffusion coefficient in regions of microfluidic networks of controlled geometry using the FRAP (Fluorescence Recovery After Photobleaching) technique. The geometry of the networks was based on Voronoi tessellations, and had varying characteristic length scale and porosity. For a fixed network, FRAP experiments were performed in regions of increasing size. Our results indicate that the boundary of the bleached region, and in particular the cumulative area of the channels that connect the bleached region to the rest of the network, are important in the measured value of the effective diffusion coefficient. We found that the statistical geometrical variations between different regions of the network decrease with the size of the bleached region as a power law, meaning that the statistical error of effective medium approximations decrease with the size of the studied medium, although no characteristic length scale could be defined over which the porous medium is equivalent to an effective medium.
\end{abstract}

\maketitle


\section{\label{sec:Introduction}Introduction}

Diffusion in porous media has been studied in several biological contexts, such as protein transport inside the cell~\cite{Cole1996, Dayel1999}, signaling in the brain~\cite{Rusakov1998a, Nicholson2001}, drug delivery in tumors~\cite{El-Kareh1993}, and transport of ions in muscles~\cite{McLennan1956, Safford1977}. It is acknowledged that diffusion is slowed down in porous media both due to impenetrable obstacles that reduce the available space for diffusion and to the intricate trajectories that particles need to describe in order to circumvent the obstacles. The first effect can be quantified in terms of the porosity of the medium, $\phi$. The second effect is usually described in terms of tortuosity, $\lambda > 1$, which quantifies the increase in path length that diffusive particles travel within the medium. Hence, when diffusion of some species is measured in the porous medium filled with a given fluid, an average, or effective diffusion coefficient, $D_\textrm{eff}$, is obtained, which is smaller than the diffusion coefficient that would be measured for the same species in the same fluid but in a free space, $D_0$.

Typical experimental values of $D_\textrm{eff}/D_0$ are $\sim 0.44$ for packed beds of beads~\cite{Delgado2008}, $\sim 1$ for the brain extracellular space~\cite{Nicholson2001}, and between 0.23 and 0.59 for the endoplasmic reticulum~\cite{Sbalzarini2005}. In real biological situations, however, one has to deal with complex processes that further slow down the diffusive transport, such as a usually unknown viscosity of the ambient fluid, binding of the diffusive molecules to macromolecular complexes, hindering to diffusion due to molecular crowding~\cite{Saxton1990, Dlugosz2011} and geometrical blockage of large diffusive molecules in narrow passages~\cite{Saxton1993}, and permeating boundaries~\cite{Latour1994}. In order to asses the relevance of these effects, it would be desirable to isolate them from the purely geometrical ones of the medium.

Theoretical models describing the porous medium as the space around a loosely packed bed of solid spheres~\cite{Lanfrey2010}, or a collection of twisted capillaries~\cite{Mathias1983}, yield values of $D_\textrm{eff}/D_0$ of the same order of magnitude than experiments. However, these models have poor predictive power when applied to real cases, since it is not clear if they capture the essential geometrical features of a particular medium. Numerical simulations have delivered useful information~\cite{Olveczky1998}, but due to practical limitations they have usually simplified the studies to periodic lattices of polyhedral obstacles~\cite{Tao2004}, which are inherently anisotropic and unrealistic. Effective medium approximations have been developed to describe the effective diffusion coefficient of porous media based on Voronoi and Delaunay tessellations~\cite{Vrettos1989, Vrettos1990, Sahimi1997}. Effective medium approximations imply a description over a length scale much larger than any characteristic geometrical length scale of the medium. In this way, local geometrical details are averaged and only global features emerge. However, measurements and simulations of diffusion in porous media are performed in samples of finite size or in reduced regions within a sample, and hence the result can reflect both the effective properties of the medium and/or local details of the measurement position~\cite{Deng2014}. The correct description of a porous medium as an effective one depends on the correct determination of a minimum length scale over which the medium can be described as an effective one. We explore on this idea by studying diffusion in networks of channels at different length scales compared to the characteristic scale of the networks. For this, we propose an experimental approach in which two-dimensional networks of microchannels are fabricated from Voronoi tessellations to simulate random, globally isotropic porous media. We measure the diffusion of fluorescein in water, a small molecule for which geometrical hindering is negligible in the micrometric channels used, as well as binding and permeation through the walls. The diffusive properties of the networks are measured locally in different regions of the sample with the fluorescence recovery after photobleaching technique~\cite{Peters1974, Jacobson1976, Lippincott-Schwartz2001, Reits2001}, which yields a diffusion coefficient in a region of defined size $R$. By varying $R$ we can focus on the local geometrical details of the networks (small $R$) or effective properties of the whole sample (large $R$).

Our results suggest that dispersion on the measured $D_\textrm{eff}$ decrease with $R$ as a power law, and hence no inherent length scale could be defined for our random networks based on Voronoi tessellations over which the effective medium is assured to be valid. More precisely, this does not precludes the definition of an effective diffusion coefficient for each network. Instead, this means that the measured diffusion coefficient will vary from place to place, around a well-defined mean value, with a dispersion that decreases as a power law with the size of the studied regions.

The paper is organized as follows. In Sec.~\ref{sec:Experimental} we describe the design and fabrication of the random networks, and the experimental procedure to find the diffusion coefficient $D_{\text{eff}}$, based on the fluorescence recovery after photobleaching technique~\cite{Reits2001, Lippincott-Schwartz2001}. In Sec.~\ref{sec:Results} we present the results, which are discussed in Sec.~\ref{sec:Discussion}. Section~\ref{sec:Conclusions} summarizes our main conclusions.


\section{\label{sec:Experimental}Materials and methods}

\subsection{Design of networks}

Random networks of channels were designed and fabricated~\cite{Wu2012}. Briefly, Voronoi polygons were generated from a set of $n$ random seed points inside a square region of sides \SI{15}{\milli\meter}. The edges of the Voronoi polygons, which become the microchannels after the fabrication process, were then assigned with a given width $w$ between \SI{20}{\micro\meter} and \SI{125}{\micro\meter}. The disorder of the network can be adjusted from the initial set of seed points, by imposing a minimum distance $d$ between any pair of points. This is quantified by the parameter $\alpha = d/d_0$, where $d_0 = \sqrt{2 A_0/(n\sqrt{3})}$ is the distance between the seed points in a completely regular network (a honeycomb network), $A_0 = (\SI{15}{\milli\meter})^2$ being the area of the square region~\cite{Zhu2001}. Hence, $\alpha = 1$ corresponds to a completely regular honeycomb lattice, while disorder increases as $\alpha$ decreases. In our study, we arbitrarily fixed $\alpha = 0.3$.

\begin{figure}[h]
\includegraphics[width=\columnwidth]{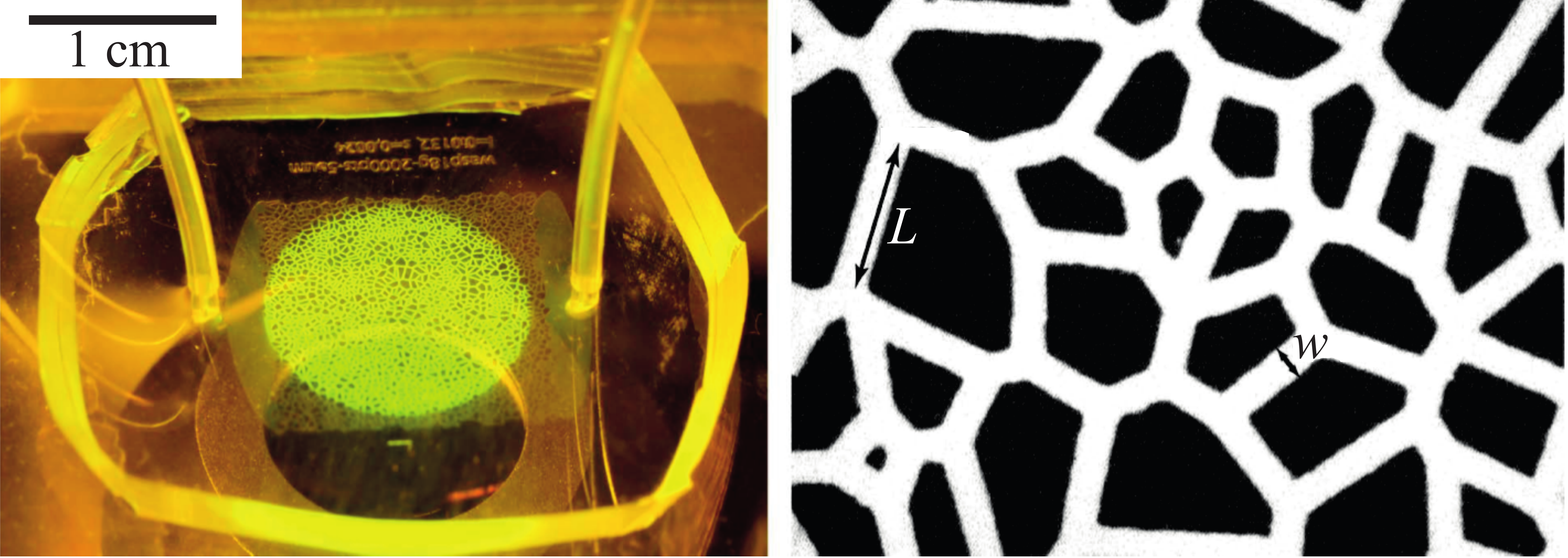}
\caption{\label{fig:Channel}Left: Photography of an assembled microfluidic channel. Inlet and outlet tubes are visible. Right: Fluorescence micrography of a region of one microfluidic network. White areas correspond to channels, filled with the fluorescent solution.}
\end{figure}

Inlet and outlet sections were added at opposite sides of the network, and then the designs were used to fabricate negative masters of the microchannels through optical lithography in SU-8 photoresist (GM1070, Gersteltech Sarl). The height of the molds was set at \SI{50}{\micro\meter}. From the masters, microfluidic networks were fabricated in polydimethylsiloxane (PDMS) (Sylgard 184, Dow Corning) using standard soft lithography techniques~\cite{McDonald2000}. Inlet and outlet access holes were punched before sealing the PDMS networks against a glass slide by oxygen plasma activation. An example of a microfluidic network is shown in Fig.~\ref{fig:Channel}.

\begin{figure}[h]
\includegraphics[width=\columnwidth]{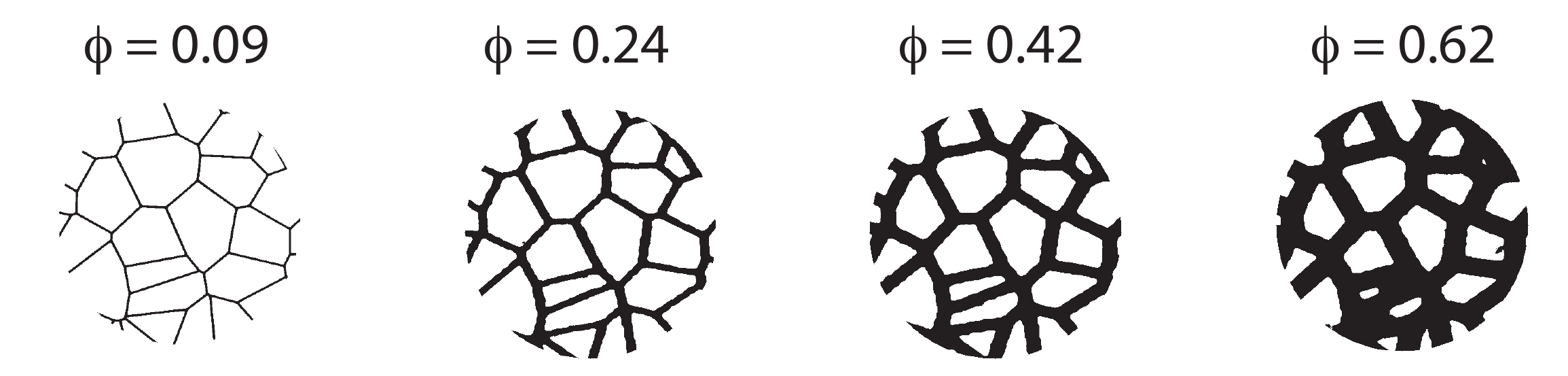}
\caption{\label{fig:Porosity}Micrographies of networks with different porosities (networks B-E from Table~\ref{tab:Channels}). As the width of the channels increases, the reticular aspect of the network is lost. Color is inverted, so dark regions represent fluorescent-filled channels.}
\end{figure}

For each network we can control the number $n$ of seed points, which determines an average length channel $\langle L \rangle$. Independently, we can control the width $w$ of the channels. Thicker channel width increases the porosity $\phi$ of the network, which we define as usual,
\begin{equation}
\phi = \frac{A_{\rm channel}}{A_0},
\label{eq:Phi}
\end{equation}
where $A_{\rm channel}$ is the total area of the microchannels in the network. As $\phi$ increases, the network resembles less a collection of narrow channels, as shown in Fig.~\ref{fig:Porosity}.

Several designs with different number of seed points and different channel widths, hence different $\phi$, were fabricated. All the channels used in this study are listed in Table~\ref{tab:Channels}.

\begin{table}[h]
\begin{tabular}{P{1.3cm}P{1.3cm}P{1.3cm}P{1.3cm}P{1.3cm}}
Network & $n$ & $\langle L \rangle$ (\si{\micro\meter}) & $w$ (\si{\micro\meter}) & $\phi$\\
\hline
\hline
A & 500 & 458 & 82 & 0.22\\
\hline
B & \multirow{4}{*}{2000} & \multirow{4}{*}{227} & 13 & 0.09\\
C & & & 38 & 0.24\\
D & & & 75 & 0.42\\
E & & & 125 & 0.62\\
\hline
F & 10000 & 101 & 24 & 0.25
\end{tabular}
\caption{\label{tab:Channels}Networks used within this study.}
\end{table}

Besides the random networks, one circular chamber of diameter \SI{1.5}{\centi\meter} and height \SI{50}{\micro\meter} was fabricated with SU-8 walls on a glass slide, filled with the fluorescent solution and closed with a cover slip in order to measure diffusion in free, two dimensional space. Finally, a ``flow-focusing'' microchannel was used to measure the diffusion coefficient in flow by measuring the widening of a central fluorescent solution sheathed by two water streams~\cite{Culbertson2002}.

\subsection{FRAP experiments}

After fabrication, the networks were filled with an aqueous solution of fluorescein (Kingscote Chemicals) at a concentration of \SI{0.5}{\milli\molar}. Microchannels were mounted on an inverted microscope (Nikon TS100) with epifluorescence illumination, and observed with a CMOS camera (DCC1545M, Thorlabs) at a rate between 1 and 1/5 frames per second (fps).

The effective diffusion coefficient of fluorescein was measured through the fluorescence recovery after photobleaching (FRAP) technique. The technique consists in irreversibly inactivating (bleaching) the fluorescence of fluorescein molecules with a high intensity excitation illumination in a small region. Afterwards, fluorescence is observed with a lower intensity excitation illumination. As bleached molecules diffuse out of, and non-bleached fluorescent molecules diffuse into the bleached area, the fluorescence is recovered in the the bleached region. Fluorescence recovery curves are then fitted with a model based on the diffusion equation~\cite{Axelrod1976, Soumpasis1983} and an effective diffusion coefficient is obtained. In our experiments, bleaching was achieved using the highest intensity of the epifluorescence excitation illumination, focused into the sample with a 10X, 0.25 NA objective during \SI{5}{\second}. The radius $R$ of the bleached area was controlled with a diaphragm between \SI{250}{\micro\meter} and \SI{1.1}{\milli\meter}. Note that the time duration of the bleaching is considerably smaller than the diffusion time scale over a region of size $R$, as the diffusion coefficient of fluorescein in water at room temperature (which varied between \SI{20}{\celsius} and \SI{25}{\celsius}) was measured at $D_0 = \SI{6.66e-10}{\meter}^2/\si{\second}$ (see Sec.~\ref{sec:D0}) and the time scale for diffusion in a two-dimensional circle of radius $R$ is given by
\begin{equation}
\tau = \frac{R^2}{4D_0},
\label{eq:Tau}
\end{equation}
which ranges between $\tau = \SI{23}{\second}$ and $\tau = \SI{454}{\second}$ for the given values of $R$. 

Recovery of the fluorescence was observed through a 2X, 0.06 NA objective using a dimmed excitation illumination to minimize further bleaching (1/32 times the intensity used for bleaching), during a period of time that ranged between \SIlist{15;60}{\min}, depending on $R$.

\begin{figure}[h]
\includegraphics[width=\columnwidth]{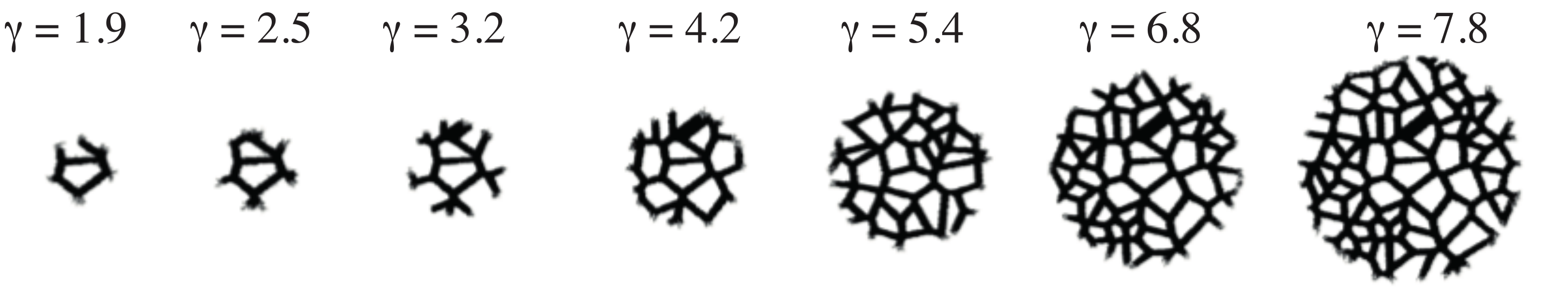}
\caption{\label{fig:Homogeneity}Color-inverted micrographies of bleached areas with different $\gamma$. Images correspond to network E in Table~\ref{tab:Channels}.}
\end{figure}

Depending on the radius of the bleached zone, more or less details of the network are included in the bleached area. To quantify the effect of geometrical details on the recovery of fluorescence, we define the homogeneity parameter
\begin{equation}
\gamma = \frac R{\langle L \rangle}.
\label{eq:Gamma}
\end{equation}
A value of $\gamma < 1$ indicates that only a fraction of a straight channel is bleached. On the other hand, $\gamma \gg 1$ means that many straight segments of the networks are included in the bleached area, and geometric details of the network are averaged in the bleached region. Examples of different bleached areas with various values of $\gamma$ are shown in Fig.~\ref{fig:Homogeneity}.

\subsection{Recovery curve and curve fitting}\label{sec:Method}

\begin{figure}[h]
\includegraphics[width=\columnwidth]{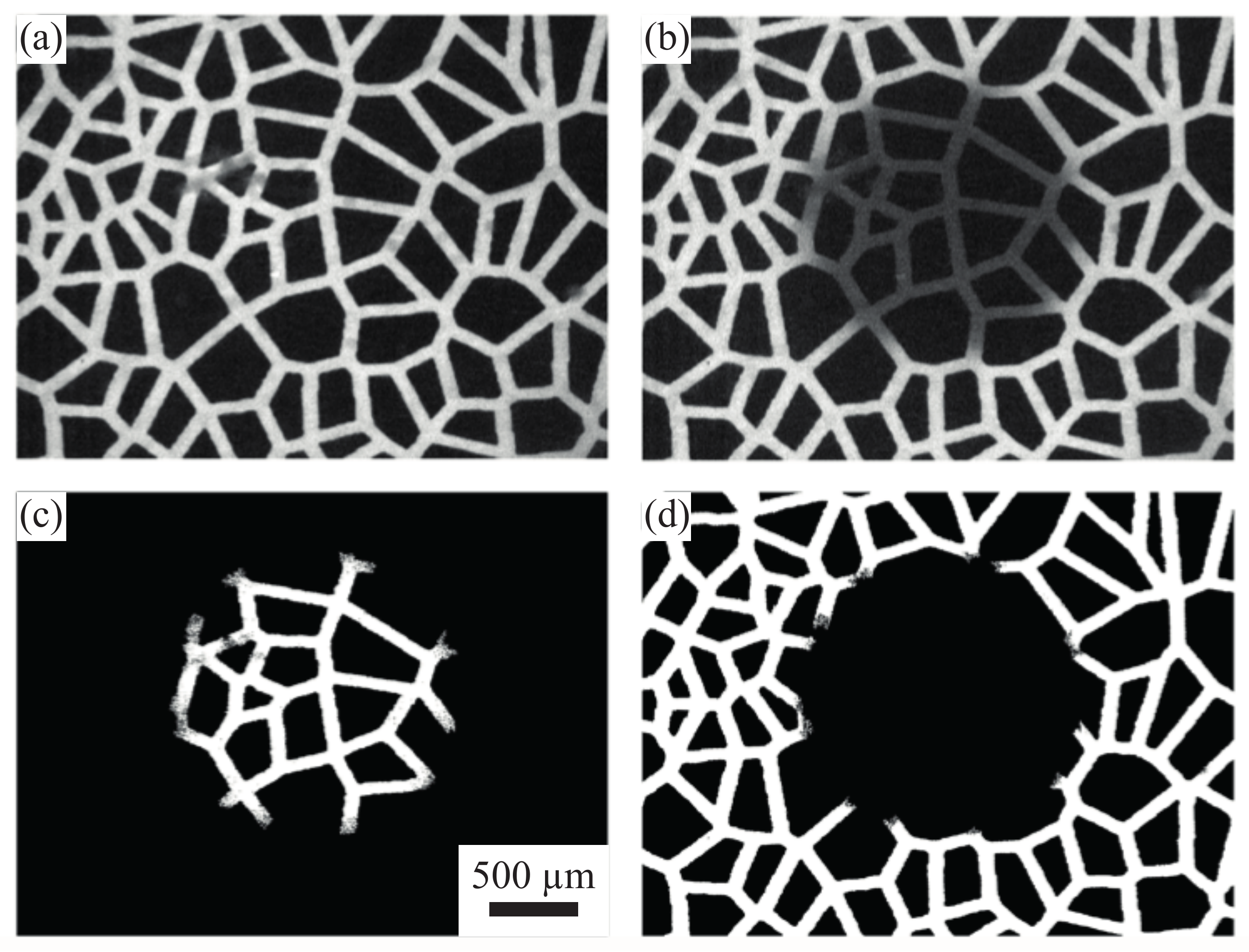}
\caption{\label{fig:Masks}Procedure for obtaining the bleached and complementary masks. (a) Image before photobleaching. (b) First image after photobleaching. (c) and (d) Bleached and complementary masks.}
\end{figure}

Recovery curves were obtained in the following manner. Prior to the experiment, a pre-bleaching image of the network was recorded with the 2X objective and low excitation intensity. A binary mask $M$ of the bleached region was obtained by dividing the pre-bleached image and the first post-bleaching image, then binarizing with application of a threshold (see Figs.~\ref{fig:Masks}(a)-(c)). A complementary mask, $M^c$ was also obtained by subtracting the mask $M$ to the binarized pre-bleaching image, as shown in Fig.~\ref{fig:Masks}(d). The average fluorescence in the bleached and complementary regions were then computed as
\begin{eqnarray}
F_\textrm{in} &= \frac{\sum_{i,j} I_{ij}(t) M_{ij}}{\sum_{i,j} M_{ij}},\\
F_\textrm{out} &= \frac{\sum_{i,j} I_{ij}(t) M^c_{ij}}{\sum_{i,j} M^c_{ij}},
\end{eqnarray}
where $I_{ij}(t)$ represents the gray value of the pixel $(i, j)$ of the post-bleaching image $I$ at time $t$, and $M_{ij}$ and $M^c_{ij}$ represent the value of the masks $M$ and $M^c$ at the pixel $(i, j)$.

\begin{figure}[h]
\includegraphics[width=\columnwidth]{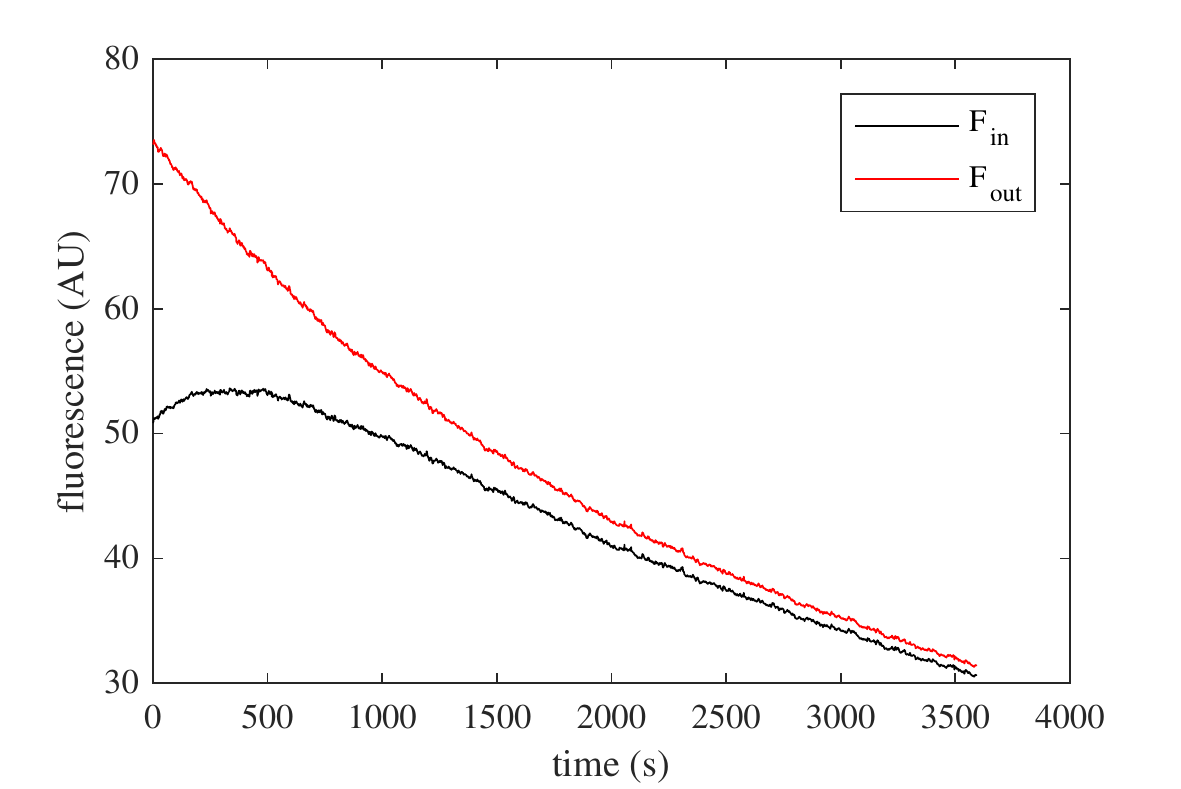}
\caption{\label{fig:ObservationBleaching}Evolution of the average fluorescence of the bleached and complementary regions.}
\end{figure}

As fluorescent molecules diffuse into, and bleached molecules diffuse out of the bleached area, the average fluorescence of the bleached area, $F_\textrm{in}$ increases and the average fluorescence of the complementary region, $F_\textrm{out}$, decreases. However, unwanted, inhomogeneous, non-negligible bleaching occurs during observation of the fluorescence recovery, causing $F_\textrm{in}(t)$ not to increase monotonically, and $F_\textrm{out}(t)$ to decrease both due to diffusion and bleaching, as shown in Fig.~\ref{fig:ObservationBleaching}. To correct for this, images of the same region of the microfluidic network are obtained with the same illumination and acquisition protocol, but without the initial bleaching. In this way, only the bleaching due to observation is recorded as a reference. The average fluorescence of the bleached and complementary regions for these reference images are computed in the same way for these reference images, as
\begin{eqnarray}
F^\textrm{ref}_\textrm{in} &= \frac{\sum_{i,j} I^\textrm{ref}_{ij}(t) M_{ij}}{\sum_{i,j} M_{ij}},\\
F^\textrm{ref}_\textrm{out} &= \frac{\sum_{i,j} I^\textrm{ref}_{ij}(t) M^c_{ij}}{\sum_{i,j} M^c_{ij}},
\end{eqnarray}
where $I^\textrm{ref}_{ij}$ represents the gray value of pixel $(i, j)$ of the reference image $I^\textrm{ref}$ at time $t$. Hence, the fraction of fluorescence recovered due to diffusion in the bleached area with respect to the complementary region is defined as
\begin{equation}
F(t) = \frac{F_\textrm{in}(t)/F_\textrm{in}^\textrm{ref}(t)}{F_\textrm{out}(t)/F_\textrm{out}^\textrm{ref}(t)}.
\end{equation}

Effective diffusion coefficients are obtained by adjusting $F(t)$ to a numerically obtained model recovery curve, similar to what is described in Ref.~\cite{Axelrod1976} but taking into account the intensity profile for bleaching. The recovery curve $F(t)$ is not normalized between 0 and 1, as is usually done, in order to increase the sensitivity of the fit. The fitting procedure has two fitting parameters, the amount of bleaching, $G$, and the characteristic timescale
\begin{equation}
\tau = \frac{R^2}{4D_\textrm{eff}},
\end{equation}
from which the effective diffusion coefficient, $D_\textrm{eff}$, is obtained.


\section{\label{sec:Results}Results}

\subsection{Absolute diffusion coefficient of fluorescein} \label{sec:D0}

The absolute diffusion coefficient of fluorescein in water, $D_0$, was measured in two different ways. First, FRAP measurements were performed in the circular quasi-two-dimensional chamber. Two different bleaching radii were used, $R_1 = \SI{198}{\micro\meter}$ and $R_2 = \SI{567}{\micro\meter}$. The same numerical fitting procedure used in the random networks was employed here, yielding the values of $D_0^{(1)} = (6.77 \pm 0.23) \times 10^{-10}\si{\meter^2/\second}$, $D_0^{(2)} = (6.35 \pm 0.25) \times 10^{-10}\si{\meter^2/\second}$, respectively. Errors represent the standard deviation of five repetitions, and are below 5\%.

On the other hand, the diffusion coefficient was determined in the flow-focusing microchannel~\cite{Culbertson2002}. Briefly, a central stream of the fluorescent solution and two flanking water streams were injected at constant flow rates. For a fixed location of the microchannel, the fluorescence intensity profile of the central stream was obtained and fitted to a Gaussian curve. The width of the Gaussian curve increases linearly with the downstream location, from which the diffusion coefficient was obtained at $D_0^{(3)} = (6.87 \pm 0.61) \times 10^{-10}\si{\meter^2/\second}$. The error corresponds to the standard deviation of seven measurements performed with different flow rates.

All results thus obtained are close to each other. In average, the free-space diffusion coefficient for fluorescein in water was considered to be $D_0 = (6.66 \pm 0.69) \times 10^{-10}\si{\meter^2/\second} $.

\subsection{Number of feeding channels}\label{sec:Feeding}

\begin{figure}[h]
\includegraphics[width=\columnwidth]{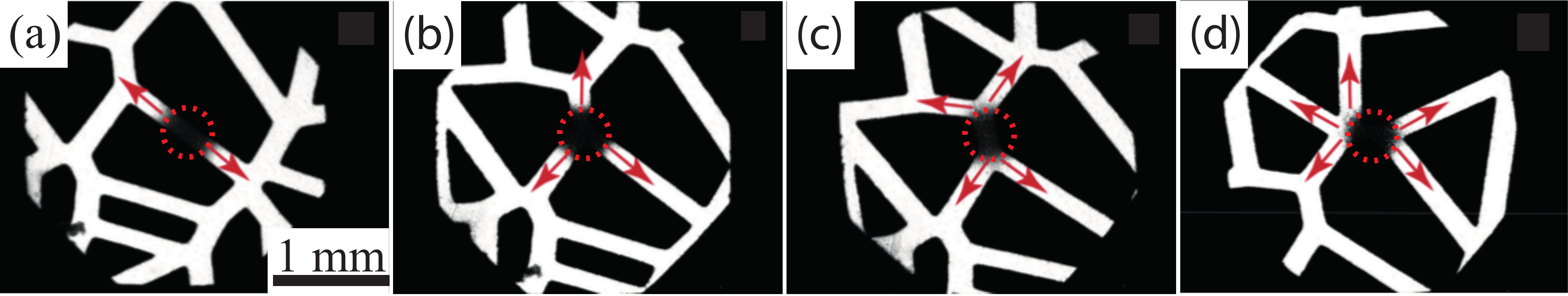}
\includegraphics[width=\columnwidth]{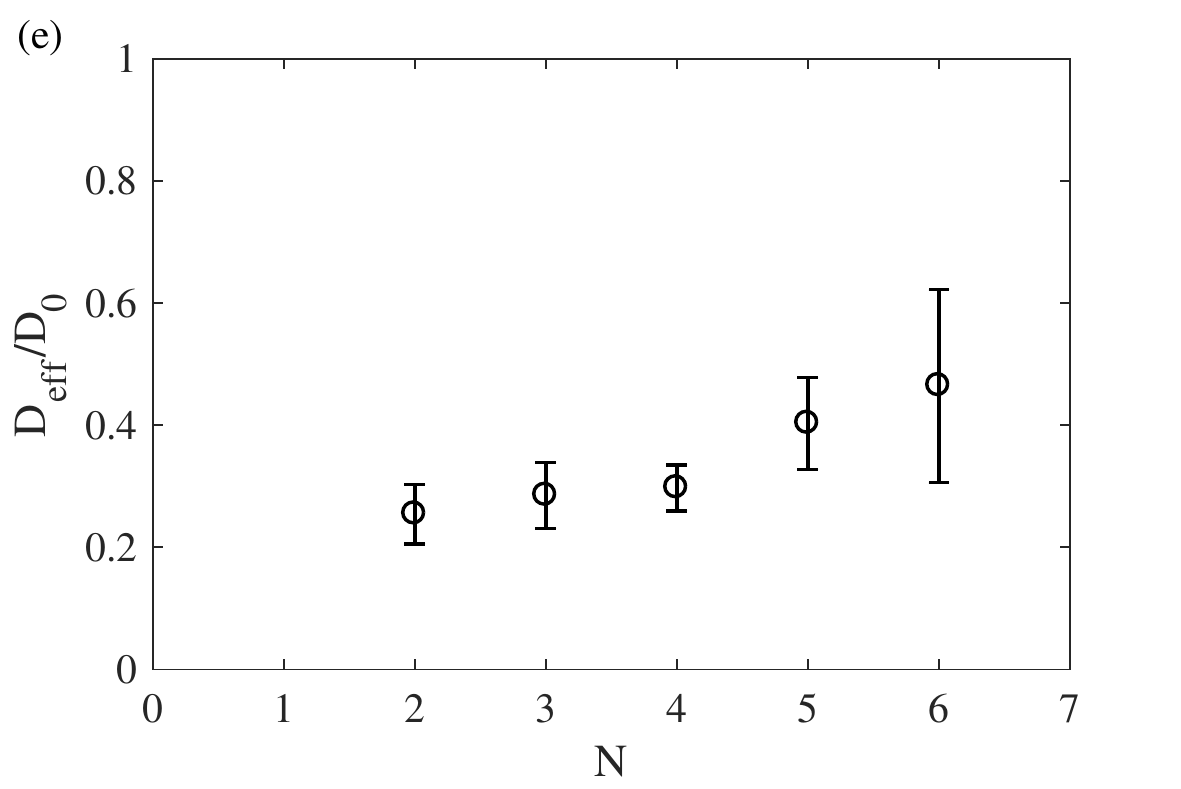}
\caption{\label{fig:FeedingChannels}(a)-(d) Images of FRAP experiments with varying number of feeding channels. All cases correspond to network A in Table~\ref{tab:Channels}. Dashed red circles indicate the bleached region. (e) Normalized effective diffusion coefficient, $D_\textrm{eff}/D_0$ for different number of feeding channels, $N$.}
\end{figure}

We investigate the effective diffusion coefficient in small bleaching regions located at the intersection between channels, as shown in Fig.~\ref{fig:FeedingChannels}(a)-(d). The objective is to establish the role that the number of channels that converge into the bleaching region, $N$, has in the recovery process. These experiments were performed in network A with a bleaching radius $R = \SI{251}{\micro\meter}$, which is smaller than the average length of the network's channels ($\gamma = 0.6$).

For each number $N$ of converging channels, five regions were randomly selected in the network. The average effective diffusion coefficient, $D_\textrm{eff}$, normalized by $D_0$ is plotted as a function of the number of feeding channels in Fig.~\ref{fig:FeedingChannels}(e), with error bars representing the standard deviation of the five measurements. The effective diffusion coefficient increases with the number of feeding channels, from a value of $\approx 0.25 D_0$ for $N = 2$ to $\approx 0.45 D_0$ when $N = 6$.

\subsection{Effective medium} \label{sec:EffectiveMedium}

In order to determine the minimum radius $R$ for which the network can be considered as an effective medium, we perform FRAP experiments using a single network (network F from Tab.~\ref{tab:Channels}) and we vary the size of the bleached region, from $\gamma = 1.33$ ($R = \SI{134}{\micro\meter}$) to $\gamma = 13.59$ ($R = \SI{1373}{\micro\meter}$). For each value of $\gamma$, ten FRAP experiments were performed in random locations of the network to obtain an effective diffusion coefficient $D_\textrm{eff}$ and its statistical variation $\Delta D_\textrm{eff}$ from the average and standard deviation of the obtained coefficients.

\begin{figure}[h]
\includegraphics[width=\columnwidth]{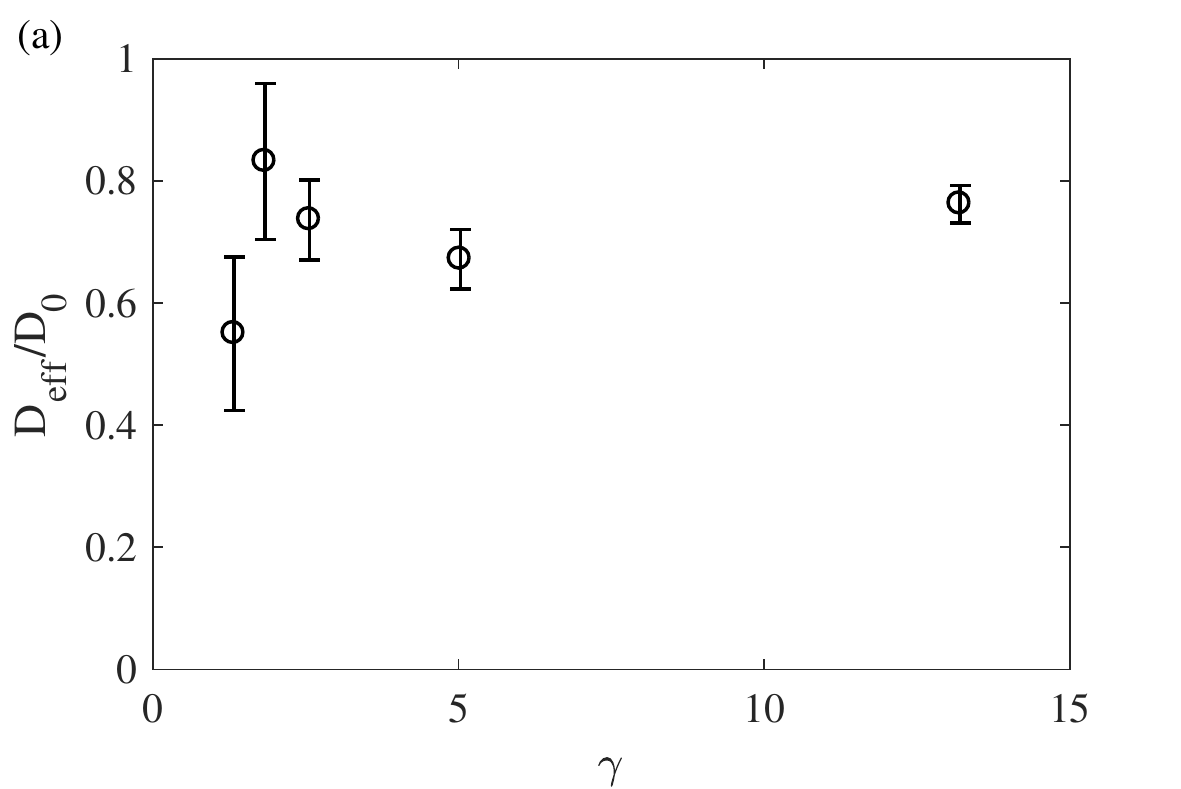}\\
\includegraphics[width=\columnwidth]{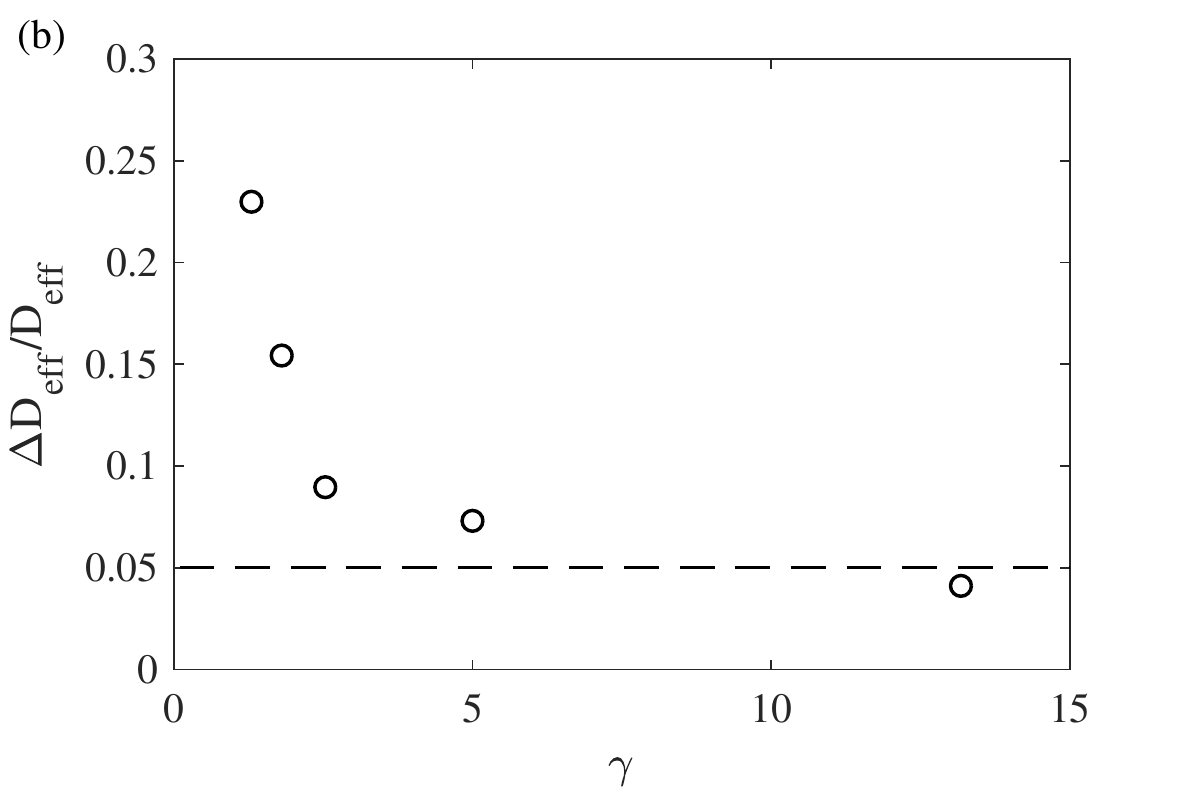}
\caption{\label{fig:EffectiveMedium}(a) Normalized effective diffusion coefficient, $D_\textrm{eff}/D_0$ as a function of $\gamma$ for network F. (b) Standard deviation of $D_\textrm{eff}$, normalized by $D_\textrm{eff}$, for ten realizations in arbitrary locations of the same network, at fixed $\gamma$. The dashed line marks the 5\% accuracy limit of our measurements.}
\end{figure}

The results are shown in Fig.~\ref{fig:EffectiveMedium}. In all cases, $D_\textrm{eff}/D_0 < 1$, indicating that the geometry of the network slows down the diffusion coefficient, as expected. The average values of $D_\textrm{eff}$ vary between $0.55 D_0$ and $0.62 D_0$ in a non monotonic way as $\gamma$ is increased in a decade (Fig.~\ref{fig:EffectiveMedium}(a)).

The dependence of the fractional error $\Delta D_\textrm{eff}/D_\textrm{eff}$ with $\gamma$ is shown in Fig.~\ref{fig:EffectiveMedium}(b). For small bleached area, the dispersion in the fitted diffusion coefficient is larger than 20\%. As the size of the bleached region increases, $\Delta D_\textrm{eff}$ rapidly decreases. For $\gamma \gtrsim 10$, $\Delta D_\textrm{eff}$ becomes comparable to the characteristic error of the method, which is estimated at $\sim5\%$ (dashed line in Fig.~\ref{fig:EffectiveMedium}(b)), according to the measurements for $D_0$ presented in Section~\ref{sec:D0}. 

\subsection{Available area}

Finally, in order to determine the effect of the available area for diffusion, we perform experiments in networks with the same geometry but varying channel width (networks B-E from Tab.~\ref{tab:Channels}). In these networks, the original Voronoi polygons are the same, and only the channel width is varied. In all cases, we fix the size of the bleaching region at $\gamma = 3.99$ ($R = \SI{790}{\micro\meter}$). Six locations were randomly chosen in the geometry to perform the FRAP experiments. We uses the same six bleaching positions in all networks, as can be noted in Fig.~\ref{fig:Porosity}, although minor displacements occurred between different networks.

\begin{figure}[h]
\includegraphics[width=\columnwidth]{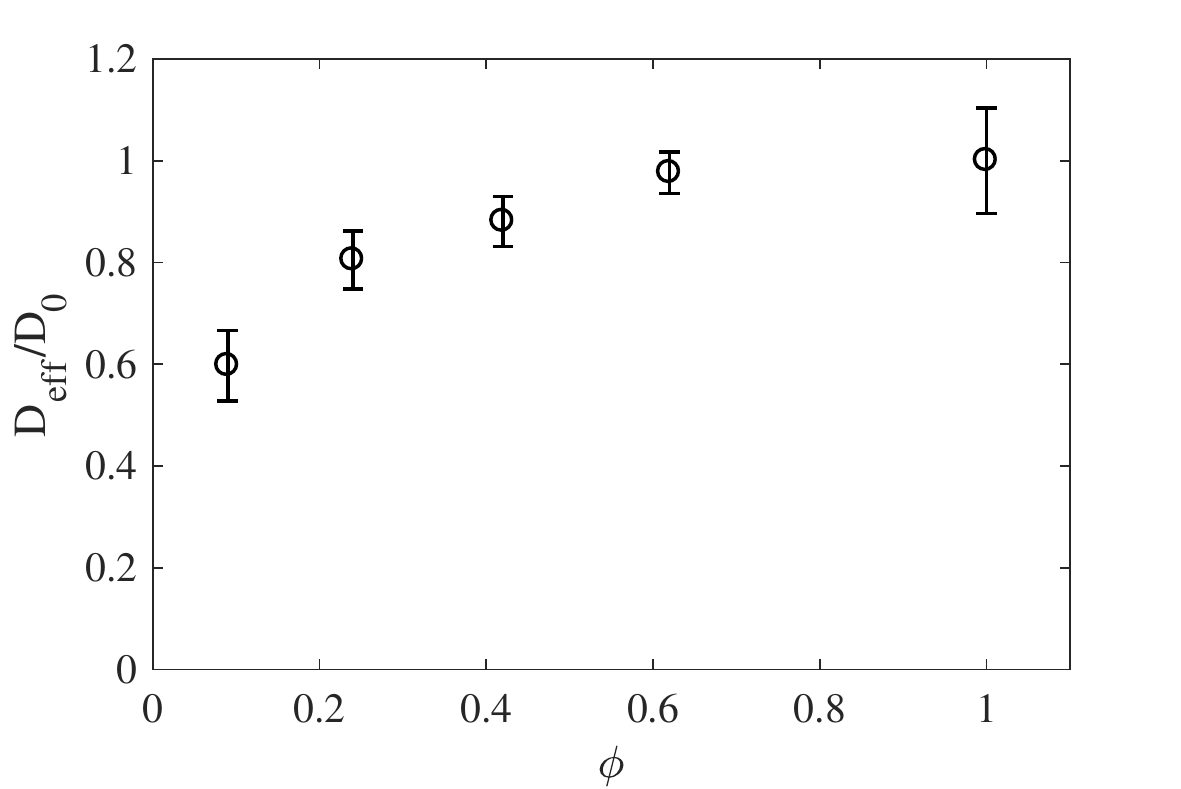}
\caption{\label{fig:AvailableArea}Normalized effective diffusion coefficient, $D_\textrm{eff}/D_0$ as a function of the porosity $\phi$ for fixed $\gamma = 3.99$.}
\end{figure}

The results are shown in Fig.~\ref{fig:AvailableArea}. The point $\phi = 1$ corresponds to a hypothetical network for which channels are so wide that merge into a quasi-two-dimensional chamber, in which case the absolute diffusion coefficient $D_0$ should be recovered. The error bar in this case corresponds to the propagated error of the three different measured values of $D_0$ (section~\ref{sec:D0}). As channels become more slender and $\phi$ decreases, the normalized effective diffusion coefficient decreases. In these cases, the error bars represent the standard deviation in the $D_\textrm{eff}$ obtained in all six locations for each network. Note that, for the chosen size of the bleaching region, the dispersion in the results between different regions is expected to be higher than the precision of the method (5\%), which is consistent with the values obtained for the error bars (between 5\% and 11\%).

\section{\label{sec:Discussion}Model for determination of characteristic length scale}

FRAP experiments in small regions of our random networks of channels, presented in Sec.~\ref{sec:Feeding}, demonstrate the relevance of the number of channels that feed the bleached region with the diffusing species. On the other hand, the dependence of $D_\textrm{eff}$ with the size of the bleached region was studied in Sec.~\ref{sec:EffectiveMedium}, in an attempt to establish the scale at which the random network can be approximated as an effective medium. In order to relate these two results, we present here a model based on the diffusion equation and on the statistical characteristics of the geometries used to fabricate the networks.

Consider, similarly to the experiments, the region consisting of the intersection between a two-dimensional random network and a circle of radius $R$, to which we call $\Omega$. Assuming a uniform excitation illumination, the fluorescence recorded within $\Omega$ is given by
\begin{equation}
F_\Omega(t) = q \int_\Omega c(r,t) d^2r,
\end{equation}
where $c(r,t)$ is the concentration of non-bleached fluorescein molecules and $q$ is a constant that accounts for all efficiencies of absorption, emission and detection of fluorescence and the excitation illumination intensity, which we will omit in the following.

Neglecting the unwanted bleaching due to observation that occurs in the experiments, the rate of change of fluorescence recorded within the region is given by the incoming flux of fluorescent molecules into $\Omega$. In comparison with free space, where the diffusive flux comes across the whole circumference of radius $R$, in the random network the incoming flux comes only through the $N$ feeding channels that intersect the circumference,
\begin{equation}
\dot F_\Omega (t) = \sum_{i=1}^N J_i.
\end{equation}
$J_i$ is given by Fick's first law integrated in the corresponding portion of the circumference, of perimeter $\mathcal P_i$, and oriented at an angle $\theta_i$,
\begin{equation}
J_i = D_0 \int_{\mathcal P_i} \nabla c \cdot \hat r d\ell \approx D_0 \left. \frac{\partial c}{\partial r} \right|_{(R,\theta_i, t)} \mathcal P_i 
\end{equation}

Assuming isotropy, we write
\begin{eqnarray}
\dot F_\Omega (t) &= D_0 \left. \frac{\partial c}{\partial r} \right|_{(R, t)} \sum_{i=1}^N \mathcal P_i\\
&= D_\textrm{eff} 2\pi R \left. \frac{\partial c}{\partial r} \right|_{(R, t)}, \label{eq:Effective}
\end{eqnarray}
where we can identify the effective diffusion coefficient as
\begin{equation}
D_\textrm{eff} = \frac{D_0}{2\pi R} \sum_{i=1}^N \mathcal P_i.
\end{equation}
This suggests that an important feature of a patterned space in its diffusing properties, at least as measured with FRAP experiments, is the fraction of open area that transports the diffusing species into the bleached region. This picture is incomplete, since the concentration field in the effective medium, say $c_\textrm{eff}$, is different to the actual concentration of the species, $c$, and should appear in (\ref{eq:Effective}). However, if we assume that the argument is qualitatively correct, it yields to an interesting conclusion regarding the definition of a length scale appropriate to define an effective medium, at least in channels based of Voronoi polygons like ours, as we shall present next.

\begin{figure}[h]
\includegraphics[width=\columnwidth]{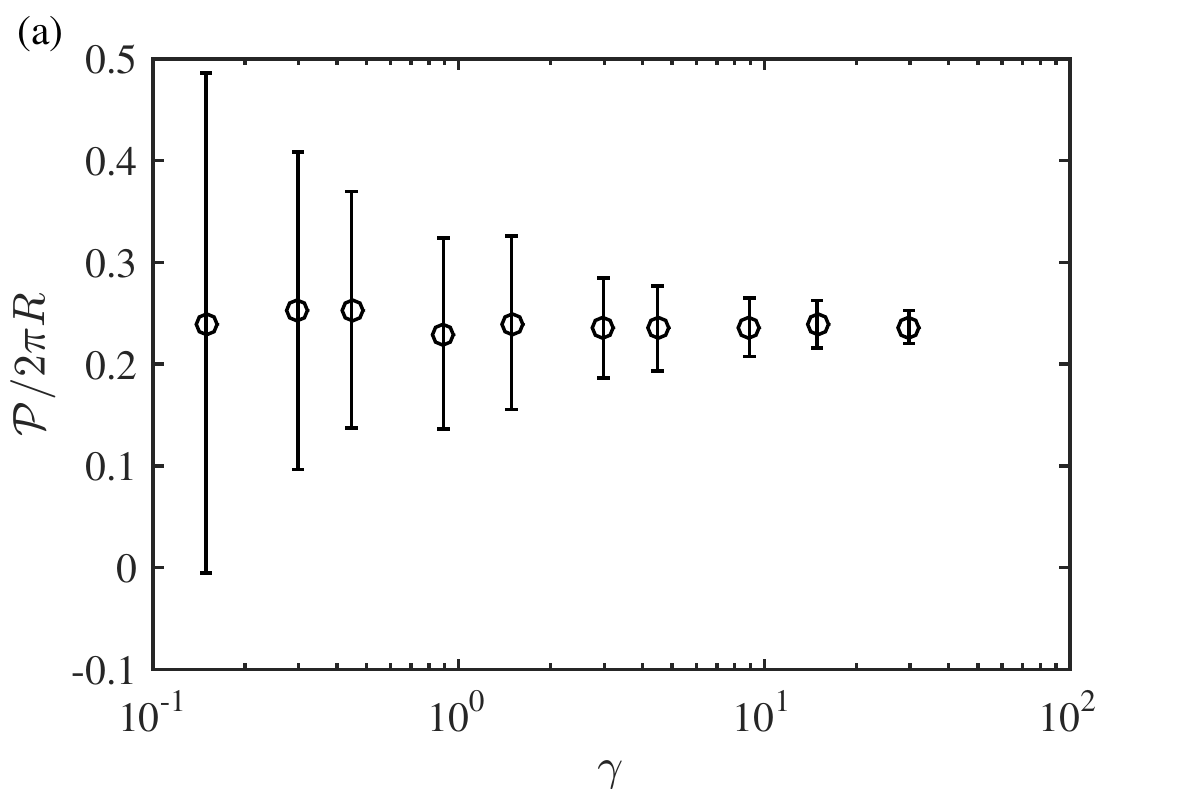}
\includegraphics[width=\columnwidth]{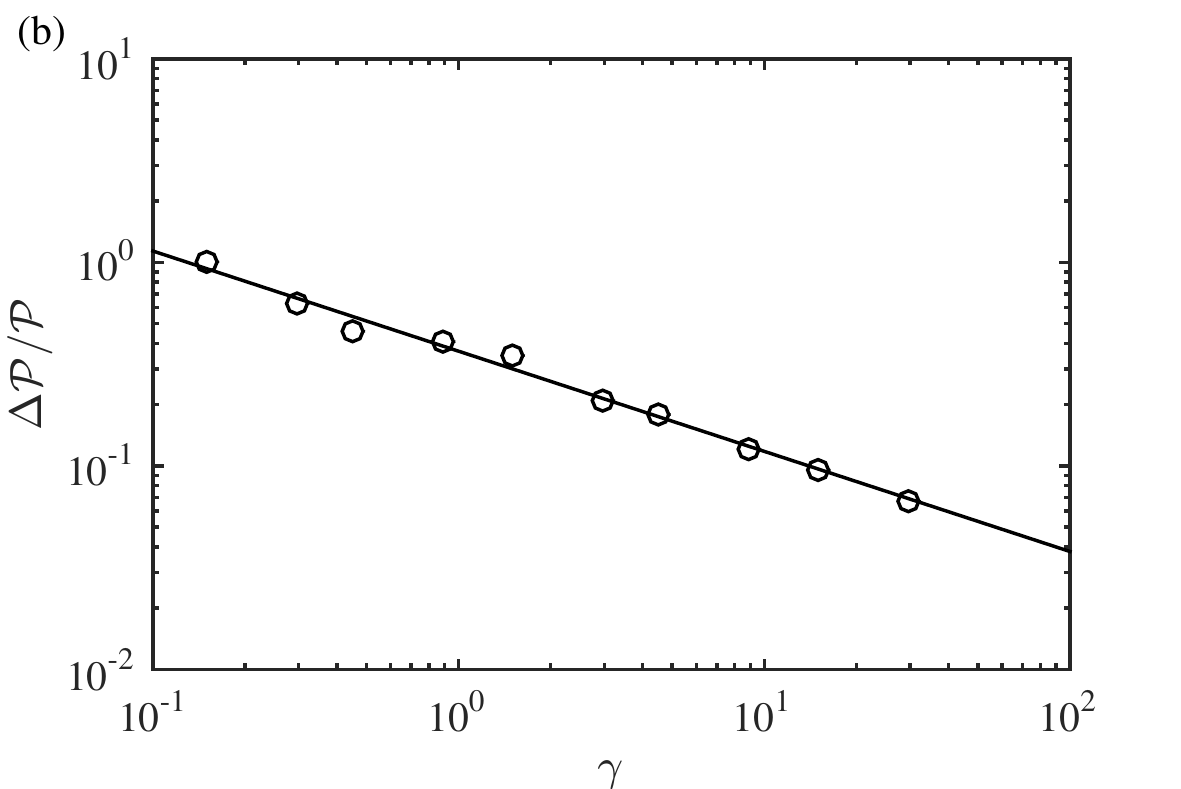}
\caption{\label{fig:ArtificialNetwork}(a) Perimeter fraction of feeding channels as a function of $\gamma$. (b) Normalized standard deviation of the perimeter fraction of feeding channels as a function of $\gamma$. The solid line represents the fit of Eq.~(\ref{eq:PowerLaw}).}
\end{figure}

As illustration, we consider the geometry of network H in Table~\ref{tab:Channels}. All initial Voronoi polygons were stored in a Matlab routine and shrunk to produce the simulated network of \SI{24}{\micro\meter}-wide channels. At 121 equally spaced locations, circles of increasing radius $R$ were drawn and the total perimeter of feeding channels, $\mathcal P = \sum \mathcal P_i$, was computed for each location and radius. Results are presented in Fig.~\ref{fig:ArtificialNetwork} as a function of $\gamma$, which is defined, as in the experiments, as $\gamma = R/\langle L \rangle$. The average perimeter fraction, $\mathcal P/2\pi R$, is shown in Fig.~\ref{fig:ArtificialNetwork}(a). Little variation occurs in the whole investigated range of $\gamma$, which spans over two decades. Error bars represent the standard deviation of the perimeter fraction, $\Delta \mathcal P/2\pi R$, which decrease dramatically as $\gamma$ increases. Figure~\ref{fig:ArtificialNetwork}(b) presents the normalized standard deviation of $\mathcal P$. As demonstrated by the linear trend in the log-log plot, $\Delta \mathcal P/\mathcal P$ follows a power law with $\gamma$, which was fitted as
\begin{equation} \label{eq:PowerLaw}
\frac{\Delta \mathcal P}{\mathcal P} = 0.34 \ \gamma^{-0.49}.
\end{equation}

The normalized standard deviation, $\Delta \mathcal P/\mathcal P$, is important to determine whether or not the network can be considered as an effective medium at the scale $\gamma$. Here, $\Delta \mathcal P/\mathcal P$ quantifies the effects of geometrical variation of inlet area between the different regions of the random network, and thus, for a truly effective medium in which the location of measurement is irrelevant, $\Delta \mathcal P/P$ should be zero. Our results indicate that $\Delta \mathcal P/\mathcal P$ decreases with $\gamma$, ie., that the variations in geometry between different regions decrease as the size of the considered region increases. Although the power law found excludes the existence of a characteristic length scale over which the effective medium approximation can be used, our results do indicate that the statistical differences between regions can be made as small as wanted by considering sufficiently large regions. Typically, one should aim to decrease the dispersion to the accuracy limit of the method, which was estimated at 5\% in our case. Experimentally, for our Voronoi-based random networks, this limit was reached for bleached regions of radius approximately 10 times larger than the mean cell length (Fig.~\ref{fig:EffectiveMedium}(b)).


\section{\label{sec:Conclusions}Discussion and Conclusions}

We have presented here an experimental approach to investigate the relevance of the geometry of porous media to diffusion properties measured with the FRAP technique. Our experiments are performed in controlled microfluidic environments, whose geometry is based on Voronoi polygons. However, other two-dimensional geometries can be used as well, such as assemblies of circular obstacles or regular polygons. This could be interesting in order to determine the relevance of geometrical features that are not present in the Voronoi tessellations, such as dead ends, large accumulation chambers, or large-scale anisotropy. Moreover, a similar microfluidic approach could complement recent efforts in the numerical reconstruction of cellular structures~\cite{Sbalzarini2005, Cui-Wang2012}. Indeed, microfluidic networks of virtually any two-dimensional geometry can be fabricated, which can aid in determining the role of geometry in the diffusive transport. Unfortunately, however, no standardized methods exist yet to fabricate complex three-dimensional microfluidic geometries, limiting the applicability of this idea to two-dimensional networks.

The method used in this work to generate the porous media, based on Voronoi polygons, can be used to vary independently the whole geometry of the network (ie. its tortuosity) by changing the seed points used to generate the polygons, or its porosity, by changing the width of the channels. Keeping constant the tortuosity of the networks demonstrates the relevance of the porosity in the effective diffusion coefficient (networks B-E and Fig.~\ref{fig:AvailableArea}). It is less evident to keep the porosity constant and vary the tortuosity in a systematic way. Comparison between networks C ($\phi = 0.24, D_\textrm{eff} = (0.80 \pm 0.06) D_0$, see Fig.~\ref{fig:AvailableArea} and H ($\phi = 0.25, D_\textrm{eff} = (0.76 \pm 0.03)D_0$, Fig.~\ref{fig:EffectiveMedium}(a)) suggests that reduction of available space, and not the particular path length, is the major geometric source of the slowing down of diffusion in porous media compared with free space. These results are consistent with existing work with aligned and staggered polygons~\cite{El-Kareh1993}.

The results and the model presented here suggest that the total area connecting the bleached and the outer region plays an important role in the effective diffusion properties measured by the FRAP technique. In particular, for small regions where the number $N$ of feeding channels can be easily counted, the effective diffusion coefficient was found to increase with $N$. For larger regions, where $N$ is not easily determined, we argue that variations of the perimeter fraction that connects with the outer region can explain in part the dispersion of the effective diffusion coefficient measured in different regions.

Applying that idea, we found that no characteristic length scale can be determined for our networks based on Voronoi polygons over which FRAP experiments are to measure a truly effective medium. Nevertheless, we found that the statistical variations between different regions of our medium decrease as a power law with the size of the bleached region, with exponent $\approx -0.5$, and thus can be made as small as the inherent accuracy limit of the FRAP method.


\begin{acknowledgments}
Useful discussions with Jorge Toledo are gratefully appreciated. MLC acknowledges support of FONDECYT grant No. 1170411 and Millenium Nucleus Physics of Active Matter of the Millenium Scientific Initiative of the Ministry of Economy, Development and Tourism (Chile). Molds for the microfluidic networks were fabricated in the Laboratory of Optical Lithography, built thanks to FONDEQUIP grants EQM140055 and EQM180009. WP acknowledges funding from Secretar\'ia de Educaci\'on Superior, Ciencia y Tecnolog\'ia del Ecuador.
\end{acknowledgments}


\end{document}